\documentclass[letterpaper,journal]{IEEEtran}
\usepackage{amsmath,amsfonts}
\usepackage{algorithm}
\usepackage{array}
\usepackage[caption=false,font=scriptsize,labelfont=sf,textfont=sf]{subfig}

\usepackage{textcomp}
\usepackage{stfloats}
\usepackage{url}
\usepackage{verbatim}
\usepackage{graphicx}
\usepackage{xcolor}
\usepackage{cite}
\usepackage{algpseudocode} 
\usepackage{booktabs}       
\usepackage{orcidlink}
\usepackage{balance}

\begin{document}

\title{Online Energy Management for Bidirectional EV Charging with Rooftop PV: An Aging-Aware MPC Approach}

\author{Francesco Popolizio\,\orcidlink{0009-0003-3719-8137},{~\IEEEmembership{Student Member,~IEEE}}, Albert Škegro\,\orcidlink{0000-0003-3819-2215},{~\IEEEmembership{Student Member,~IEEE}}, Torsten Wik\,\orcidlink{0000-0002-5234-8426},\\{~\IEEEmembership{Member,~IEEE}}, Chih Feng Lee\,\orcidlink{0000-0003-4337-3723},{~\IEEEmembership{Senior Member,~IEEE}}, and Changfu Zou\,\orcidlink{/0000-0001-7119-6854},{~\IEEEmembership{Senior Member,~IEEE}}
\thanks{This work was supported by the Swedish Energy Agency within the Vehicle Strategic Research and Innovation Program (Grant No.~P2022-00960) and the European Union's Horizon Europe program under the Marie Skłodowska-Curie (Grant No.~101131278).}
\thanks{F. Popolizio, A. Škegro, T. Wik and C. Zou are with the Department of Electrical Engineering, Chalmers University of Technology, Göteborg, Sweden
        ({\tt\footnotesize frapop@chalmers.se, skegro@chalmers.se, torsten.wik@chalmers.se, changfu.zou@chalmers.se})}%
\thanks{Chih Feng Lee is with Polestar Performance AB, Göteborg, Sweden
        ({\tt\footnotesize chih.feng.lee@polestar.com})}%
}



\maketitle

\begin{abstract}
This paper investigates the economic impact of vehicle-home-grid integration in the presence of rooftop PV, by proposing an online, aging-aware energy management strategy for an electric vehicle (EV), a household, and the electrical grid. The model predictive control-based framework explicitly exploits vehicle-to-grid (V2G) and vehicle-to-home (V2H) operation to perform energy arbitrage, increase self-consumption, while respecting user-driven driving requirements. The framework optimizes power flows over a shrinking horizon using a detailed battery aging model that captures both calendar and cycle degradation, and a Transformer-based forecaster that provides short-term predictions of household load and solar irradiance.
For a one-year horizon, the proposed strategy yields the lowest annual cost among all evaluated strategies. Adding PV increases the annual profit by €1060.7 compared to operating without PV, and yields an economic gain of up to €2410.5 over smart unidirectional charging, at the expense of only 1.27\% extra battery degradation. Even in the least favorable case with no remuneration for V2G energy, bidirectional operation still delivers an economic gain of €355.8 through V2H. 
Sensitivity analyses over V2G price ratio, EV battery size, household demand, and pickup time uncertainty confirm that these benefits persist across a wide range of scenarios and highlight the potential of EVs as active energy nodes, enabling sustainable energy management and cost-effective battery usage in real-world conditions.

\end{abstract}


\section{Introduction}
\IEEEPARstart{I}{n} the past decade, the adoption of electric vehicles (EVs) has accelerated rapidly, driven by decarbonization policies and technological progress in battery systems.
During 2025, sales of EVs are expected to have surpassed 20 million \cite{2025IEAGlobalEVOutlook}.
This widespread adoption brings clear benefits in terms of reduced local emissions and lower reliance on fossil fuels, but it also increases electricity demand in the transport sector \cite{2025IEAGlobalEVOutlook}.
Such growth introduces new challenges for power systems, particularly in ensuring adequacy, reliability, and sustainability of electricity supply.
In parallel with the growth of EVs, the deployment of photovoltaic (PV) systems has also been expanding rapidly worldwide \cite{IEA2024WorldEnergyInvestment}. 
However, the variable and weather-dependent nature of solar generation add further variability and uncertainty to the grid. 
Addressing these challenges requires flexibility that can balance demand and supply at different timescales, which EVs can potentially provide through bidirectional charging technologies.

With the emergence of bidirectional power transfer for EVs, there is potential to reduce costs for their owners while also providing services to the grid. In this context, concepts such as vehicle-to-grid (V2G) and vehicle-to-home (V2H) are increasingly adopted, where EVs are no longer merely means of transportation but can also act as energy nodes. 
V2G enables EVs to supply energy back to the grid, offering services such as load balancing and frequency regulation, while allowing the EV owners to generate profit by selling stored energy. V2H allows EVs to supply power to a home, supporting home energy management and enabling the user to reduce energy costs and increase self-sufficiency \cite{2019v2xReview}.

Despite this potential, the practical value of bidirectional charging is ultimately determined by both technical feasibility (e.g., mobility constraints and EV battery aging) and economic viability under realistic tariffs and operating conditions.
Therefore, rigorous techno-economic assessment and energy management design are needed to quantify when V2G and V2H, with PV coordination, can deliver tangible benefits.

In the relevant literature, most approaches rely on offline optimization. 
For example, the authors of \cite{2022v2hv2gGermany} explored V2G and V2H applications for energy trading, reporting that an average German household with a PV system and a stationary battery can generate annual revenues of €310. 
A mixed-integer linear programming based home energy management framework was proposed in \cite{2015v2hv2g}, integrating PV generation, bidirectional EV utilization (V2H and V2G), and a stationary battery, with the objective of minimizing the daily electricity cost for households.
Without considering V2H, \cite{2023MARLv2g} formulated the day-ahead V2G scheduling problem within the multi-agent reinforcement learning framework to optimize the peak shaving performance for the electric grid.
In \cite{2019JapanEVv2hPV}, PV generation supplies the household load, and surplus energy is sold to the grid, while an EV and a stationary battery are considered to support the household. The authors show that, for non-commuter users, V2H is more advantageous than a stationary battery.
A similar conclusion is reported in \cite{2025PVoptimalsizingAustralia}, where V2H is argued to replace stationary batteries while avoiding additional costs and still meeting household demand.
A common limitation of the aforementioned works is that they assume full knowledge of input data in advance, such as household loads and PV generation, and then solve the problem offline, making them less applicable to real-world scenarios.

To overcome these drawbacks, recent research has increasingly focused on online optimization and control strategies, which can adapt to uncertainties and changing conditions in real time.
In \cite{2019EVchargingPV}, a four-stage control algorithm was developed for an EV charging station with PV generation and stationary storage integrated with a commercial building.
An online battery anti-aging V2G scheduling method based on fuzzy logic control was proposed in \cite{2022onlineV2GPengfei} to solve the optimization in two stages; however, only cycle aging was considered in the offline calibration used to tune the fuzzy control parameters.
In \cite{2024OnlineV2GPSO}, particle swarm optimization was used to solve the V2G optimization with a sliding window; as for the battery degradation, it only considers the cycle aging with the rain-flow cycle counting method.
More recently, \cite{2025NNMPCv2gPV} introduced a neural network-based model predictive control scheme for a grid-connected PV system with a stationary battery and an EV, enabling V2G coordination but neglecting battery degradation.

Despite these advances, existing online optimization methods rely on overly simplified battery degradation models, often accounting only for cycle aging while neglecting calendar effects. 
Furthermore, the integration of V2G, V2H, battery dynamics, PV generation, and household demand under uncertainty has not yet been fully investigated, and none of the aforementioned models considers range anxiety, i.e., the need to preserve sufficient EV battery charge to guarantee mobility, which is a key factor in the practical adoption of bidirectional charging strategies.
These limitations highlight the need for more competent and realistic frameworks capable of real-time adaptability while mitigating the impact on battery lifespan.

To address the identified research gaps, this paper proposes an online control algorithm for a single-user system comprising a house, an EV, and a rooftop PV installation.
The framework aims to minimize user costs, including both electricity costs and battery degradation, while V2G operations are specifically exploited for energy arbitrage, i.e., purchasing electricity during low-price hours and selling during high-price hours.
A detailed nonlinear battery aging model, accounting for calendar and cycle aging under coupled stress factors, is implemented and subsequently linearized for tractable integration into the optimization framework.
In addition, PV generation is incorporated as a flexible source, capable of supplying the household demand, charging the EV, or exporting surplus power to the grid. 
While electricity prices are known in advance from the day-ahead market, household load and PV generation are uncertain and only measured up to the current time. A Transformer-based neural network is therefore used to forecast future household load and solar irradiance (and hence PV generation), and these forecasts are updated at each time step using newly available measurements so that the controller always optimizes based on the most recent information.
Range anxiety is also addressed by enforcing EV state-of-energy (SoE) constraints that guarantee sufficient mobility for the user. 
As a result, the proposed algorithm minimizes the overall user electricity and battery costs, while contributing to more sustainable and economically viable energy management.

\section{Vehicle-home-grid-pv Control}
\label{sec:2}
\begin{figure}
    \centering
    \includegraphics[width=0.90\linewidth]{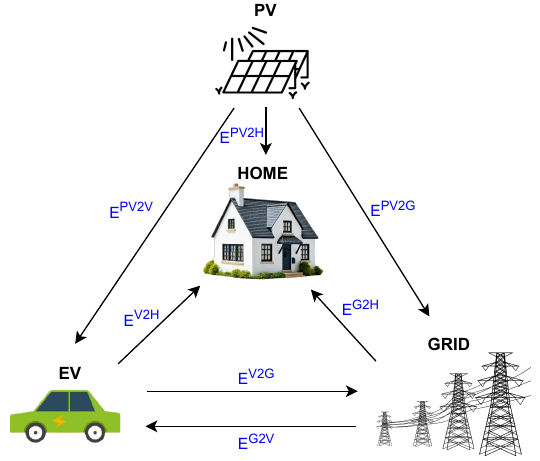}
    \caption{Vehicle-home-grid-PV integration in the considered system. The arrows indicate the allowable energy flows for V2G, V2H, G2V, G2H, PV2V, PV2H and PV2G.}
    \label{fig:scen}
\end{figure}

The considered system, as illustrated in Fig. \ref{fig:scen}, has four actors: an EV, a house, a PV system, and the power grid. 
The EV can supply energy to the grid and the house (V2G and V2H), while grid-to-vehicle (G2V) represents the energy used to recharge the EV. The house, on the other hand, can also receive energy directly from the power grid through grid-to-home (G2H). The PV system can supply energy to the household demand (PV2H), charge the EV (PV2V), and export any surplus to the grid (PV2G).
The present work assumes that the EV is only charged when parked at home.

The proposed framework considers both driving and parking phases. When the EV is driving, no optimization is performed, as the EV simply follows the user's driving profile, while the house and PV operate normally. When parked, all actors interact, allowing optimal energy management within the system.

The proposed control strategy is formulated as a shrinking-horizon model predictive control (SH-MPC) problem, activated whenever the EV is parked. The prediction horizon corresponds to the parking duration, known only at the beginning of each parking session. As time evolves, the horizon shrinks until the EV is picked up, leading to an online optimization framework that continuously recalculates the optimal energy flows using updated forecasts and real-time data.

\subsection{Energy Flow Control in Parking Mode}
While the EV is parked, the SH-MPC framework computes the optimal energy flow by solving a constrained optimization problem. 
The problem is formulated in discrete time with a sampling period of $1$ hour, where $t$ indexes hourly intervals (i.e., the interval $[t,\,t+1)$) and $\bar{t}$ denotes the present time index.
Denoting $t_a$ and $t_p$ as the EV arrival and pickup hours, the optimization is solved at each $\bar{t}\in \{t_a, t_a+1, \ldots, t_p\}$ over the remaining parking horizon $t=\bar{t}, \ldots, t_p$ by minimizing the total cost of energy exchanges and battery degradation:
\begin{equation} \label{eq:obj}
    \min \sum_{t=\bar{t}}^{t_p} \mathit{EC}_t - \mathit{ER}_t + \mathit{BC}_t,
\end{equation}
where $\mathit{EC}_t$ is the energy cost, $\mathit{ER}_t$ is the energy revenue by selling energy back to the grid, and $\mathit{BC}_t$ is the battery degradation cost.

The terms $\mathit{EC}_t$, $\mathit{ER}_t$ and $\mathit{BC}_t$ can be calculated by
\begin{align}    
    \mathit{EC}_t =\:& (E^{\text{G2V}}_t + E^{\text{G2H}}_t) \cdot p_t,  
    \label{eq:EC}
    \\ 
    \mathit{ER}_t =\:& (E^{\text{V2G}}_t \cdot \gamma + E^{\text{PV2G}}_t)  \cdot p_{\mathrm{s},t},  
    \label{eq:ER}
    \\ 
    \mathit{BC}_t =\:& \mathrm{NV} \cdot \frac{\Delta Q_{\text{loss},t}(\%)}{100\% - \mathrm{EoL}(\%)}, 
\label{eq:BC}
\end{align}
where the energy flows are expressed in kWh, $p_t$ denotes the retail purchase price, including spot price, taxes, grid fees, and VAT, while $p_{\mathrm{s},t}$ is the spot price used to remunerate energy exports to the grid. 
Energy costs $\mathit{EC}_t$ arise from purchasing energy from the grid for G2V and G2H, while energy revenues $\mathit{ER}_t$ arise from selling energy from the EV and the PV to the grid (V2G and PV2G).
The parameter $\gamma \in [0,1]$, denoted as price ratio, is a V2G remuneration factor that scales the V2G export term in \eqref{eq:ER}, so that energy sold via V2G is paid $\gamma \cdot p_{\mathrm{s},t}$. As $\gamma$ decreases, the V2G selling price decreases, reducing the economic incentive to export energy via V2G.
In \eqref{eq:BC}, $\Delta Q_{\text{loss},t}$ is the battery degradation (described in Section~\ref{sec:battery}) and $\mathrm{EoL}$ (End of Life) battery capacity is assumed to be 80\% of the nominal value. $\mathrm{NV}$ is the battery net present value and, based on the economic principles in \cite{park2019_economics}, is calculated as follows:
\begin{equation}    \label{eq:NV}
    \mathrm{NV} = \frac{C_{\mathrm{rep}}(L) - C_{\mathrm{rv}}(L)}{(1+i_r)^L},
\end{equation}
where $L$ denotes the nominal battery life in years, $C_{\mathrm{rep}}(L)$ and $C_{\mathrm{rv}}(L)$ are, respectively, the battery replacement cost and residual value evaluated after $L$ years, and $i_r$ is the yearly discount rate, i.e., the time value of money, used for present-value discounting.

The objective function in \eqref{eq:obj} is subject to several constraints that ensure feasibility and consistency of the energy flows.
In particular, the following constraints apply to the energy flows and the limits of both the EV charger and the PV system:
\begin{align}    
    E^{\text{V2G}}_t, E^{\text{V2H}}_t, E^{\text{G2V}}_t, E^{\text{G2H}}_t, & \nonumber \\
    E^{\text{PV2V}}_t, E^{\text{PV2G}}_t, E^{\text{PV2H}}_t & \geq \: 0, \label{eq:c1}
\end{align}
\begin{equation}
    E^{\text{G2V}}_t + E^{\text{PV2V}}_t \leq \: E_{\mathrm{EV},\max}, \label{eq:c2}
\end{equation}
\begin{equation}
    E^{\text{V2G}}_t + E^{\text{V2H}}_t \leq \: E_{\mathrm{EV},\max}, \label{eq:c3}
\end{equation}
\begin{equation}
    E^{PV2G}_t + E^{PV2H}_t + E^{PV2V}_t \leq\: E_{PV,\max}.    \label{eq:c4}
\end{equation}
In \eqref{eq:c1}, all energy flows must be non-negative. The input energy ($E^{\text{G2V}}_t+E^{\text{PV2V}}_t$) and the output energy from the EV ($E^{\text{V2G}}_t+E^{\text{V2H}}_t$) cannot exceed the maximum limit $E_{\mathrm{EV},\max}$, dictated by the charging interface, leading to \eqref{eq:c2}-\eqref{eq:c3}. 
Finally, \eqref{eq:c4} restricts the maximum energy that can be exported by the PV system, according to the rated limit imposed by the inverter.

The state of energy (SoE) of the EV battery is subject to constraints that enforce admissible SoE bounds, battery dynamics, and conditions related to the pickup requirement:
\begin{equation}
    10\% \leq \mathrm{SoE}_t \leq \: 90\%, \label{eq:c5}
\end{equation}
\begin{equation}
    \mathrm{SoE}_t = \mathrm{SoE}_{t-1} + \frac{E^{\text{G2V}}_t + E^{\text{PV2V}}_t}{E_{\mathrm{b}}} - \frac{E^{\text{V2G}}_t + E^{\text{V2H}}_t}{E_{\mathrm{b}}}, \label{eq:c6}
\end{equation}
\begin{equation}
    \mathrm{SoE}_{t_p}  \geq\: \mathrm{SoE}_p,  \label{eq:c7}
\end{equation}
where, $E_{\mathrm{b}}$ denotes battery capacity, and $\mathrm{SoE}_p$ represents the minimum required SoE at the pickup hour $t_p$; the SoE bounds in \eqref{eq:c5} are set to 10\% and 90\% to avoid overcharging and deep discharging of the battery.

According to \eqref{eq:c7}, if the user retrieves the EV earlier than expected, for instance, due to sudden plan changes, the SoE could be insufficient for an unplanned trip, leading to range anxiety.
Considering a battery with a nominal range of 500 km, a 40\% SoE corresponds approximately to 200 km, which constitutes a reasonable safety margin to cover unexpected driving needs \cite{2025mpcV2Grangeanxiety}.
If the EV is parked from $t_a$ to $t_p$ (both endpoints included), the total parking duration in hours is
\begin{equation}
T_p = t_p - t_a + 1. \label{eq:Tp_def}
\end{equation}
Based on this definition, an additional constraint is introduced to enforce a minimum SoE of 40\% during the second half of the parking period:
\begin{equation}
\mathrm{SoE}_{t_a + i} \ge 40\%, \quad \forall\, i \in \left[ \left\lfloor \tfrac{T_p}{2} \right\rfloor,\, T_p - 1 \right] .  \label{eq:c8}
\end{equation}
This provides a safety margin for unexpected early pickup or unplanned trips, thereby mitigating range anxiety.

The system operation is further constrained by a set of balance equations that describe the energy exchanges between the household, the grid, and the PV system:

\begin{equation}
    \mathit{HL}_t =\: E^{\text{G2H}}_t + E^{\text{V2H}}_t + E^{\text{PV2H}}_t, \label{eq:c9}
\end{equation}
\begin{equation}
    E^{\text{G2V}}_t + E^{\text{G2H}}_t \leq\: G_t.     \label{eq:c10}
\end{equation}
Equation \eqref{eq:c9} represents the household energy balance, where the household load $\mathit{HL}_t$ is known at time $t$ and forecasted over the prediction horizon, and is supplied by the grid (G2H), the EV (V2H), and the PV system (PV2H). 
Constraint \eqref{eq:c10} limits the total energy drawn from the grid, which cannot exceed the contracted energy $G_t$, assumed sufficiently large so that the constraint is never active in this study.

The PV generation is modeled separately to explicitly show how the available solar energy is determined. In particular, the total PV energy produced at each time step is computed as
\begin{equation}       
    E^{\mathrm{PV}}_t = \mathrm{SR}_t \cdot E^{\mathrm{PV}}_{\text{tot}},     \label{eq:c11}
\end{equation}
where \(\mathrm{SR}_t \in [0,1]\) is the normalized solar irradiance profile (measured at time $t$ and forecasted over the prediction horizon), and $E^{\mathrm{PV}}_{\text{tot}}$ is the total installed PV capacity. 
Finally, the PV energy balance is given by
\begin{equation}
    E^{\mathrm{PV}}_t = E^{\text{PV2G}}_t + E^{\text{PV2H}}_t + E^{\text{PV2V}}_t + E^{\text{PV2curt}}_t,     \label{eq:c12}
\end{equation}
which ensures that the total PV production is distributed over the grid, the house, and the EV, while any surplus beyond the inverter is curtailed ($E_t^{\text{PV2curt}}$).

\subsection{Energy Flow Control in Driving Mode}
When the EV is in driving mode, it is no longer connected to the grid or the household, and therefore no optimization problem is solved.
During this phase, the SoE evolves deterministically according to the driving energy consumption, assumed to be mainly distance-driven \cite{Qingbo_EVcons}, and leading to the SoE update
\begin{equation}
\mathrm{SoE}_t = \mathrm{SoE}_{t-1}  - \frac{d_t}{\eta_{\mathrm{EV}} \cdot E_{\mathrm{b}}}  , \label{eq:driving_SoC}
\end{equation}
where $d_t$ is the distance traveled during the interval $[t,t+1)$, and $\eta_{\mathrm{EV}}$ is the EV energy efficiency, expressed in km/kWh.

In parallel, the household and PV system maintain their energy exchange under a rule-based strategy.
The PV generation first supplies the household load, while any surplus is sold to the grid.
If the PV generation is insufficient to meet the household demand, the missing portion is supplied by the grid.
This logic ensures that the household can still benefit from solar generation even when the EV is absent.

During the driving phase, the total costs are still evaluated in terms of energy cost ($\mathit{EC}$), energy revenue ($\mathit{ER}$), and battery cost ($\mathit{BC}$) as in the parking mode, but only the grid-to-home and PV-to-grid flows contribute to $\mathit{EC}$ and $\mathit{ER}$, respectively.

Battery degradation continues to accumulate while driving, as a result of both calendar and cycle aging, with the latter depending on the energy consumed during the trip.
This contribution is added to the cumulative degradation state to ensure continuity between the driving and parking phases.

\section{Battery degradation model}

\label{sec:battery}
    Accurate modeling of battery degradation is essential for optimizing bidirectional charging strategies, as these must balance grid service revenues against the cost of accelerated battery wear. 
    In this work, a semi-empirical degradation model for nickel–manganese–cobalt (NMC) batteries, capturing both calendar and cycle aging under stress factors such as temperature, SoE, and depth of discharge (DoD), is adopted from~\cite{2023batteryNature}. 
    Compared to lithium iron phosphate (LFP), NMC batteries exhibit stronger sensitivity of degradation to cycling conditions, rendering them particularly relevant for degradation-aware optimization of bidirectional charging strategies.

    \subsection{Semi-Empirical Degradation Model}
    
        Battery degradation is modeled as the battery capacity loss. The total battery capacity loss $Q_{\text{loss}}$ is modeled as the sum of calendar- and cycle-induced degradation:
        \begin{equation}
            Q_{\text{loss}} = Q_{\text{loss}}^{\text{cal}} + Q_{\text{loss}}^{\text{cyc}}.
            \label{eq:q_total}
        \end{equation}        

        The calendar aging model follows a square-root dependence on time:
        \begin{equation}
            Q_{\text{loss}}^{\text{cal}} = K^{\mathrm{nlin}}_{\mathrm{cal}} \cdot \sqrt{t_{\text{stor}}},
            \label{eq:q_cal}
        \end{equation}
        where
        \begin{equation}
            t_{\text{stor}} = (t+1)\Delta t,
            \label{eq:t_stor}
        \end{equation}
        with $\Delta t = 1~\mathrm{h}$, denotes the elapsed storage time in hours. The calendar aging rate $K^{\mathrm{nlin}}_{\mathrm{cal}}$ is modeled using an Arrhenius-type expression that captures the temperature dependence through an activation energy term and the SoE dependence through the anode open-circuit potential $U_{\mathrm{a}}$:
        \begin{align}
            K^{\mathrm{nlin}}_{\mathrm{cal}} = K^{\mathrm{ref}}_{\mathrm{cal}} & \cdot \exp\left( -\frac{E_{\mathrm{a}}}{R} \left( \frac{1}{T} - \frac{1}{T_{\mathrm{ref}}} \right) \right) \nonumber \\ 
            & \cdot \exp\left( \frac{\alpha F}{R} \left( \frac{U_{\mathrm{a}}}{T} - \frac{U_{\mathrm{a,ref}}}{T_{\mathrm{ref}}} \right) \right),
            \label{eq:Kcal}
        \end{align}
        where $K^{\mathrm{ref}}_{\mathrm{cal}}$ is the reference calendar aging rate, $E_{\mathrm{a}}$ is the activation energy, $R$ is the universal gas constant, $F$ is the Faraday constant, $T$ is the cell temperature, $T_{\mathrm{ref}}$ is the reference temperature, $U_{\mathrm{a,ref}}$ is the reference anode potential, and $\alpha$ is a fitting parameter. 
        Assuming a constant temperature, $T = T_{\mathrm{ref}}$, makes $K^{\mathrm{nlin}}_{\mathrm{cal}}$ a function of SoE alone through the SoE-dependence of $U_{\mathrm{a}}$.

        The cycle aging model assumes a linear dependence on current throughput:
        \begin{equation}
            Q_{\text{loss}}^{\text{cyc}} = K_{\mathrm{cyc}} \cdot \mathrm{EFC},
            \label{eq:q_cyc}
        \end{equation}
        where $\mathrm{EFC}$ represents the cumulative number of equivalent full cycles (EFCs) and is computed as
       \begin{equation}
           \mathrm{EFC} = \frac{1}{2} \sum_{t=0}^{t_{\text{end}}} |\Delta \mathrm{SoE}_t|,
           \label{eq:EFC}
       \end{equation}
       where $t_{\text{end}}$ is the total number of simulation time steps, and $\Delta \mathrm{SoE}_t = \mathrm{SoE}_t - \mathrm{SoE}_{t-1}$. 
        The cycle aging rate $K_{\mathrm{cyc}}$ depends on DoD and temperature:
        \begin{align}
            K_{\mathrm{cyc}} = K^{\mathrm{ref}}_{\mathrm{cyc}} & \cdot (k_\mathrm{A} \cdot \mathrm{DoD} + k_\mathrm{B}) \nonumber \\ 
            &\cdot \left( k_\mathrm{G} \cdot (T - T_{\mathrm{ref}})^2 + k_\mathrm{H} \right),
            \label{eq:Kcyc_app}
        \end{align}
        where $K^{\mathrm{ref}}_{\mathrm{cyc}}$ is the reference cycle aging rate, and $k_\mathrm{A}$, $k_\mathrm{B}$, $k_\mathrm{G}$, $k_\mathrm{H}$ are fitting parameters. 
        Under the constant-temperature assumption ($T = T_{\mathrm{ref}}$), $K_{\mathrm{cyc}}$ becomes a function of DoD only.
        The calendar and cycle aging parameters are listed in \cite{2023batteryNature}.

    \subsection{Linearization of the Degradation Model}
    \label{subsec:linearization}

        The optimization framework presented in Section~\ref{sec:2} employs a linear programming (LP) formulation to enable efficient real-time computation. However, the calendar aging rate $K^{\mathrm{nlin}}_{\mathrm{cal}}$ exhibits a nonlinear dependence on SoE through the anode potential $U_{\mathrm{a}}$, which prevents direct integration into the LP problem. 

        To address this, the nonlinear function $K^{\mathrm{nlin}}_{\mathrm{cal}}$ is approximated using a PWL representation with breakpoints $\{\tau_0, \tau_1, \ldots, \tau_M\}$ over the SoE operating range $[0.1, 0.9]$, following \eqref{eq:c5}:
        \begin{align}
            K_{\mathrm{cal}}^{\mathrm{lin}} \approx b_0 &+ m_0 \cdot \mathrm{SoE} \nonumber \\ &+ \sum_{i=1}^{M-1} \Delta m_i \cdot \max(0, \mathrm{SoE} - \tau_i),
            \label{eq:Kcal_pwl}
        \end{align}
        where $b_0$ is the intercept, $m_0$ is the slope of the first segment, and $\Delta m_i$ represents the slope change at each interior breakpoint $\tau_i$ (the parameters listed in Table~\ref{tab:pwl_params}). The breakpoint values are chosen using secant slopes such that the approximation passes exactly through $K^{\mathrm{nlin}}_{\mathrm{cal}}$ at each $\tau_i$. 
        The resulting fit, obtained using breakpoints at $10\%$ SoE intervals, shows close agreement with a root-mean-square error of $3.3 \times 10^{-6}$.

        \begin{table}[ht]
            \centering
            \caption{PWL approximation parameters for $K_{\mathrm{cal}}^{\mathrm{lin}}(\mathrm{SoE})$.}
            \label{tab:pwl_params}
            \begin{tabular}{l @{\hspace{30pt}} l}
                \toprule
                \multicolumn{2}{l}{\text{First-segment intercept and slope}} \\
                $b_0=-6.491 \times 10^{-6}$      &  $m_0=7.613 \times 10^{-5}$  \\
                \midrule
                \multicolumn{2}{l}{\text{Interior breakpoints and slope changes}} \\
                $\tau_1 = 0.2$ & $\Delta m_1 =1.362 \times 10^{-4}$ \\
                $\tau_2 = 0.3$ & $\Delta m_2 =1.087 \times 10^{-4}$ \\
                $\tau_3 = 0.4$ & $\Delta m_3 =-1.317 \times 10^{-4}$ \\
                $\tau_4 = 0.5$ & $\Delta m_4 =-3.668 \times 10^{-5}$ \\
                $\tau_5 = 0.6$ & $\Delta m_5 =3.324 \times 10^{-4}$ \\
                $\tau_6 = 0.7$ & $\Delta m_6 =5.757 \times 10^{-4}$ \\
                $\tau_7 = 0.8$ & $\Delta m_7 =-3.912 \times 10^{-4}$ \\
                \bottomrule
            \end{tabular}
        \end{table}

    \subsection{Incremental Degradation Formulation}

        For integration into the SH-MPC framework, the degradation model is reformulated to the incremental form. 
        Starting from the continuous calendar aging model in~\eqref{eq:q_cal}, the instantaneous degradation rate is obtained by differentiation:
        \begin{equation}
            \frac{dQ_{\text{loss}}^{\text{cal}}}{dt_{\text{stor}}}
            =
            \frac{K^{\mathrm{nlin}}_{\mathrm{cal}}}{2\sqrt{t_{\text{stor}}}}.
        \end{equation}
        which is singular at $t = 0$. To avoid this singularity, the discrete approximation evaluates the degradation rate at \(t+1\) in \eqref{eq:t_stor}, treating each time index as the end of the corresponding hour. 
        Recalling the hourly discretization, the incremental calendar aging at each time step is approximated as
        \begin{equation}
            (\Delta Q_{\text{loss}}^{\text{cal}})_t
            \approx
            K_{\mathrm{cal}}^{\mathrm{lin}} \cdot \frac{1}{2\sqrt{t+1}}.
        \end{equation}
        For cycle aging, the incremental contribution at each step is
        \begin{equation}
            (\Delta Q_{\text{loss}}^{\text{cyc}})_t = K_{\mathrm{cyc}} \cdot \Delta \mathrm{EFC}_t,
            \label{eq:dq_cyc}
        \end{equation}
        where $\Delta \mathrm{EFC}_t = |\Delta \mathrm{SoE}_t| / 2$ is the incremental EFC contribution. The cycle aging coefficient $K_{\mathrm{cyc}}$ depends on the DoD, which in the incremental formulation is approximated by the difference between the maximum and minimum values of SoE.

        The total degradation at each time step is
        \begin{equation}
            \Delta Q_{\text{loss},t} = (\Delta Q_{\text{loss}}^{\text{cal}})_t + (\Delta Q_{\text{loss}}^{\text{cyc}})_t,
            \label{eq:dq_loss_total}
        \end{equation}
        which enters the objective function~\eqref{eq:obj} through the battery cost term $BC_t$ defined in~\eqref{eq:BC}. 
        The cumulative calendar and cycle degradation over the horizon $t \in \{0, 1, \ldots, t_{\text{end}}\}$ are
        \begin{align}
            Q_{\text{loss}}^{\mathrm{cal}} &= \sum_{t=0}^{t_{\text{end}}} (\Delta Q_{\text{loss}}^{\text{cal}})_t, \label{eq:Q_cal_cumulative} \\
            Q_{\text{loss}}^{\mathrm{cyc}} &= \sum_{t=0}^{t_{\text{end}}} (\Delta Q_{\text{loss}}^{\text{cyc}})_t, \label{eq:Q_cyc_cumulative}
        \end{align}
        with total degradation $Q_{\text{loss}} = Q_{\text{loss}}^{\mathrm{cal}} + Q_{\text{loss}}^{\mathrm{cyc}}$.

\section{Transformer-Based Forecasting of Household Load and Solar Irradiance Profile}
\label{sec:4}
In real-world scenarios, household load is strongly influenced by user behavior and ambient conditions, while the solar irradiance profile is mainly determined by weather. 
Their inherent variability introduces forecast uncertainty, which can impact real-time optimal energy management while the EV is parked. 
To tackle this, a Transformer-based forecasting model is developed to predict household demand and solar generation, leveraging long-range dependency modeling and improved accuracy over traditional recurrent models~\cite{2024TransformerLoadForecasting}.

The forecasting architecture follows \cite{2017TransformerGoogle, 2024TransformerLoadForecasting}, and is adapted for one-hour-ahead forecasting ($t+1$) for both household load and solar irradiance profile. To achieve this, the network is trained using a sliding window of the past 24 hours ($t-23$ to $t$), which serves as the input sequence to the Transformer.

The Transformer-based neural network was implemented using Tensorflow and Keras. The model was trained with a batch size of 64 and for 100 epochs, employing the Adam optimizer. 
The remaining settings are reported in \cite{2017TransformerGoogle}.

Four input features were selected, based on historical data:
\begin{enumerate}
\renewcommand{\labelenumi}{$f\theenumi)$}
\item household load [kWh]
\item solar irradiance profile [-]
\item day of the year (ranging from 1 to 365)
\item hour of the day (ranging from 0 to 23)
\end{enumerate}
These features were first normalized to $[0,1]$ using Min--Max scaling, and then split into training, validation, and testing subsets. Specifically, $80\%$ of the data was used for training (of which $20\%$ served as validation), while the remaining $20\%$ was reserved for testing. The combination of physical (load, irradiance) and temporal (day, hour) features enables the model to capture both periodic and contextual dependencies across seasons and time-of-day patterns.

The forecasting module provides one-hour-ahead predictions of both household load and solar irradiance profile.  
However, in the vehicle-home-grid-PV framework, multi-step forecasts are required to anticipate future conditions over the EV parking period.
To obtain multi-hour predictions, the trained neural network is applied recursively, where each newly predicted value is fed back as input to forecast the next time step, until the required horizon is reached.

\section{Simulation Data and Model Parameters}
\label{sec:5}

This section presents the parameters and data used for the simulations, including the household load and solar irradiation profile data for the forecasting model, the EV usage profiles, electricity price data, EV battery and PV specifications.

\subsubsection{Data sources for forecasting model}
The forecasting model described in Section~\ref{sec:4} uses two input data sources: household electricity load and solar irradiance profile.

For the household load, this study uses the public dataset in \cite{2021_hl_data}, which provides hourly electricity consumption profiles for single apartments in the United States. Five households from Washington State were selected due to climatic conditions broadly comparable to southern Scandinavia, yielding similar seasonal household consumption patterns. The selected profiles were concatenated into a five-year hourly load dataset, with four years used for training and one year for testing (20\% of the total data). The test year has an average daily electricity consumption of 21.6~kWh.

For the solar irradiance profile, hourly global irradiance measurements were taken from a meteorological station in Gothenburg, Sweden \cite{2021_sun_data_goteborg}. Five consecutive years (2016--2020) were selected for this study.

\subsubsection{EV driving pattern}
The EV daily driving behavior is modeled using truncated Gaussian distributions based on real-world Swedish mobility datasets \cite{2025YukiDrivingPatterns,trafikanalys2024}. Specifically, pickup time, parking start time, and daily driving distance are sampled using the mean and standard deviation reported in Table~\ref{tab:EV_pattern}, with truncation applied at the corresponding minimum and maximum bounds. The mean daily driving distance is derived from an average annual mileage of 11410~km for passenger cars in Sweden \cite{trafikanalys2024}.
The traveled distance is assumed to be linearly distributed over the travel period, implying a constant power during driving hours.

\setlength{\tabcolsep}{6pt} 

\begin{table}[h]
\centering
\caption{Truncated Gaussian parameters for EV driving patterns.}
\label{tab:EV_pattern}
\begin{tabular}{lcccc}
\toprule
\textbf{} & \text{Mean} & \text{Std. dev.} & \text{Min.} & \text{Max.} \\
\midrule 
Pickup time (h)          & 6.7     & 0.8    & 4    & 10   \\[0.02cm]
Parking start time (h)   & 17      & 1.2    & 14   & 21   \\[0.02cm]
Daily driving distance (km)    & 31.2    & 10     & 5    & 60    \\[0.02cm]
\bottomrule
\end{tabular}
\end{table}

\subsubsection{Electricity price}
Since this work focuses on the Swedish electricity market, each household in Sweden holds two distinct contracts: one with the local Distribution System Operator (DSO), and another with an electricity supplier.
In this study, Göteborg Energi \cite{goteborgenergi} is considered as both the reference electricity supplier and the local DSO.
The retail electricity purchase price $p_t$, used in \eqref{eq:EC}, is computed as:
\begin{equation}
\label{eq:price}
    p_t = (1+r_{\mathrm{VAT}}) \cdot ( p_{\mathrm{s},t} + p_\mathrm{var} + p_\mathrm{del} + p_{\mathrm{net}} )
\end{equation}
where all components are summarized in Table \ref{tab:energy_price}.
All values were originally reported in Swedish Krona (SEK) and then converted to Euro (€) using an exchange rate of 1€ = 11~SEK.

\setlength{\tabcolsep}{6pt} 
\begin{table}[h]
    \centering    
    \caption{Electricity cost components due to tax and fees \cite{goteborgenergi}.}
    \label{tab:energy_price}
    \begin{tabular}{lccc}
    \toprule
        \text{Description}    & \text{Symbol}     &  \text{Value}                     \\
        \midrule     
        
        Value Added Tax         & $r_{\mathrm{VAT}}$  &  0.25                               \\
        Spot Price              & $p_{\mathrm{s},t}$  &  \textit{from Nord Pool} (€/kWh)    \\
        Variable Fee            & $p_\mathrm{var}$      &  0.00363 €/kWh                      \\
        Delivery Fee            & $p_\mathrm{del}$      &  0.00627 €/kWh                      \\
        Network Fee (DSO)       & $p_\mathrm{net}$     &  0.0399 €/kWh                       \\
        Monthly tax             &                     &  4.09 €/month                       \\

        \bottomrule
    \end{tabular}
\end{table}
In addition, a fixed monthly tax of 4.09 €/month is applied; this component is not included in \eqref{eq:price}, as it does not depend on the energy consumption, but it is instead added to the final cost at the end of the simulation.
Finally, $p_{\mathrm{s},t}$ corresponds to the hourly spot price for Sweden (bidding zone SE3) for year 2022, provided by Nord Pool \cite{nordpool}.

\subsubsection{EV Specification} 
In accordance with a Polestar EV, we set the battery capacity $E_{\mathrm{b}}$ to 79 kWh (usable capacity); additional EV models with different battery capacities are introduced later for comparison.
The EV efficiency is set to $\eta_{\mathrm{EV}}=$ 5.9 km/kWh \cite{EVdatabase}, and the maximum energy for charging/discharging ($E_{\mathrm{EV},\max}$) is set to 11 kWh. Also, the desired SoE at the pickup hour is set to $\mathrm{SoE}_p=80\%$.

As for the EV battery, we assume a nominal lifetime of \(L = 10\) years. 
The replacement cost \(C_{\mathrm{rep}}(L)\) is defined as the projected battery pack price at the end of this nominal life, i.e., in year 2035. 
Historical Li-ion EV battery pack costs from 2008 to 2025 are taken from \cite{BattCost_data, BattCost_data_1}; a constant compound annual growth rate (CAGR), estimated over the last 10 years of available data, is then used to project the 2035 cost, yielding \(C_{\mathrm{rep}}(L) = 52.46~\)€/kWh.

The battery residual value is set to \(C_{\mathrm{rv}}(L) = 20\% \,C_{\mathrm{rep}}(L)\) and a yearly discount rate of \(i_r = 4\%\) is adopted.

\subsubsection{PV Specification}
Regarding the PV system, the economic assumption is a one-time capital cost of 1700 €/kWh, including mounting structures, wiring, and inverter costs, with a service life of 20 years \cite{IEA_PVPS_Sweden_2023}. 
In the model, this cost is annualized over the service life, translating into 85 €/kWh for a one-year simulation. 

PV generation mainly depends on solar irradiance profile, panel orientation/tilt, and cell temperature; for simplicity, a favourable tilt and a constant cell temperature of \(25^\circ\)C are assumed in this work.

For this study, we consider a total installed PV capacity of \(E^{\mathrm{PV}}_{\text{tot}} = 10~\mathrm{kWh}\)
which corresponds to an annualized PV investment cost of \(E^{\mathrm{PV}}_{\text{tot}} \cdot 85 = 850~\text{€}/\text{year}\).
This system size is representative of a typical residential PV installation for a standard single-family house in Sweden~\cite{sizePV_SE}. 
However, additional PV system configurations with different total installed capacities $E^{\mathrm{PV}}_{\text{tot}}$ are introduced later for comparison.

Moreover, the maximum hourly energy output from PV, denoted as \(E_{\mathrm{PV},\max}\), is set to 11~kWh.

\subsubsection{Simulation settings}
To match the data resolution, simulations were conducted over one year with an hourly time step. The simulation started with the EV's SoE at 60\%. 
The complete control algorithm was developed in Python, with CasADi \cite{casadi} serving as the optimization framework and the HiGHS solver applied for solving the optimization problem.

\section{Results and Discussion}
Simulation studies are carried out to evaluate the user costs associated with the integration of V2G and V2H. Two configurations are considered: Vehicle-Home-Grid-PV (VHGPV), corresponding to Fig.~\ref{fig:scen}, and Vehicle-Home-Grid (VHG), which excludes PV generation.

For each configuration, four control strategies are compared: 

\begin{itemize}
    \item \textit{Proposed strategy} -- Minimizes total user cost by optimizing bidirectional energy flows, as defined in \eqref{eq:obj}.
    \item \textit{Unidirectional Smart Charging} -- Same objective as above, but with V2G and V2H disabled ($E^{\text{V2G}}=E^{\text{V2H}}=0$). The EV charges from the grid or PV, but is not used as an active grid node.
    \item \textit{Energy-cost minimization only} -- Minimizes only net energy cost ($\min \sum_{t=\bar{t}}^{t_p} \mathit{EC}_t - \mathit{ER}_t$), ignoring battery degradation.
    \item \textit{Battery-degradation minimization only} -- Minimizes only battery degradation cost ($\min \sum_{t=\bar{t}}^{t_p} \mathit{BC}_t$), ignoring energy cost and revenue.
\end{itemize}

\subsection{Cost and Battery Degradation Analysis}

Table~\ref{tab:result_table} summarizes user costs for the four control strategies with V2G price ratio $\gamma = 1$, i.e., V2G exports paid at the full spot price, which is the most favorable V2G scenario.
The user's final cost ($\mathit{FC}$) is computed as the sum of the net energy costs ($\mathit{EC}-\mathit{ER}$) and the battery degradation cost ($\mathit{BC}$). In addition, $\mathit{FC}$ includes the monthly energy tax of €4.09 reported in Table~\ref{tab:energy_price}, which is applied to all configurations. In scenarios including PV, $\mathit{FC}$ further accounts for the fixed annualized PV investment cost, equal to €850 for \(10~\mathrm{kWh}\) PV system.

\setlength{\tabcolsep}{4pt} 
\begin{table}[ht]
\centering
\caption{User cost breakdown and battery degradation metrics for the four control strategies with V2G price ratio $\gamma = 1$, evaluated over a one-year simulation.}
\label{tab:result_table}
\begin{tabular}{lccccccc}
\toprule
                        & $\mathit{FC}$      & $\mathit{EC}$-$\mathit{ER}$       & $\mathit{BC}$      & $Q_{\text{loss}}$    & $Q_{\text{loss}}^{\mathrm{cal}}$   & $Q_{\text{loss}}^{\mathrm{cyc}}$   & $\mathrm{EFC}$    \\
                        & [€]       & [€]        & [€]       & [\%]    & [\%]         & [\%]         & [-]      \\
\midrule
\multicolumn{8}{l}{\textbf{VHG}} \\[0.05cm]
Proposed                & 20.9    & -401.8    & 373.6    & 3.34    & 1.47         & 1.87         & 243.2   \\[0.02cm]
Unidir.                 & 2381   & 2096.5    & 235.4    & 2.10    & 1.91         & 0.19         & 24.5    \\[0.02cm]
$\min \mathit{EC}$-$\mathit{ER}$          & 47.2     & -417.4    & 415.4    & 3.71    & 1.48         & 2.23         & 290.0   \\[0.02cm]
$\min \mathit{BC}$               & 2501.3   & 2229.9    & 222.4    & 1.98    & 1.79         & 0.19         & 24.8    \\[0.02cm]
\midrule
\multicolumn{8}{l}{\textbf{VHGPV}} \\[0.05cm]
Proposed                & -1039.8  & -2313.8   & 374.9    & 3.37    & 1.48         & 1.89         & 243.2   \\[0.02cm]
Unidir.                 & 1370.7   & 236.2     & 235.4    & 2.10    & 1.93         & 0.17         & 24.5    \\[0.02cm]
$\min \mathit{EC}$-$\mathit{ER}$          & -1014.6  & -2329.1   & 415.4    & 3.71    & 1.48         & 2.23         & 288.8   \\[0.02cm]
$\min \mathit{BC}$               & 1693   & 566.6     & 227.4    & 2.03    & 1.88         & 0.15         & 24.8    \\[0.02cm]
\bottomrule
\end{tabular}   
\end{table}

The proposed strategy achieves favorable performance in both configurations. The negative net energy cost indicated high revenue from arbitrage. This comes at the expense of moderate additional degradation: 3.34\% (VHG) and 3.37\% (VHGPV). After accounting for degradation cost and fixed charges, the final cost is €$20.9$ (VHG) and €$-1039.8$ (VHGPV). Hence, adding PV improves the user's annual profit by €1060.7.

In contrast, unidirectional smart charging incurs substantially higher net energy costs but uses the battery far less: the equivalent full cycles are only 24.5 for both VHG and VHGPV over one year, corresponding to 2.10\% total degradation. Despite lower degradation cost, final cost rises to €2381 and €1370.7 for VHG and VHGPV, respectively. 

Comparing the proposed and unidirectional strategies, bidirectional operation increases degradation by only 1.24–1.27\%, with approximately ten times higher battery utilization (243.2 vs.\ 24.5 EFC). The increased cycling raises $Q_{\text{loss}}^{\mathrm{cyc}}$, but lower average SoE during bidirectional operation partially offsets this through reduced calendar aging for both configurations. The net result is an economic gain of €2360.1 for VHG and €2410.5 for VHGPV.

The remaining two strategies minimize only the net energy cost and only the battery degradation cost, respectively. When degradation is ignored in the objective function, the optimizer drives aggressive battery usage, achieving slightly better net energy costs but increasing degradation to 3.71\%. 
Once $\mathit{BC}$ is included in the final cost, performance worsens compared to the proposed strategy: $\mathit{FC}$ becomes €$47.2$ (VHG) and €$-1014.6$ (VHGPV). 

Minimizing only battery degradation yields the opposite failure mode: total degradation drops to $\sim$2\%,  but net energy cost becomes large and positive, equal to €2229.9 for VHG and €566.6 for VHGPV. Consequently, this strategy yields the highest final cost among all strategies and thus represents the least attractive option from the user's economic perspective.

Overall, these results confirm that the objective must include both energy and degradation costs. The proposed strategy provides a balanced trade-off, achieving the lowest final cost with only modest additional battery wear.

\subsection{Sensitivity Analysis of V2G Price Ratio \texorpdfstring{$\gamma$}{gamma}}

\begin{figure}[ht]
    \centering
    \includegraphics[width=0.9\linewidth]{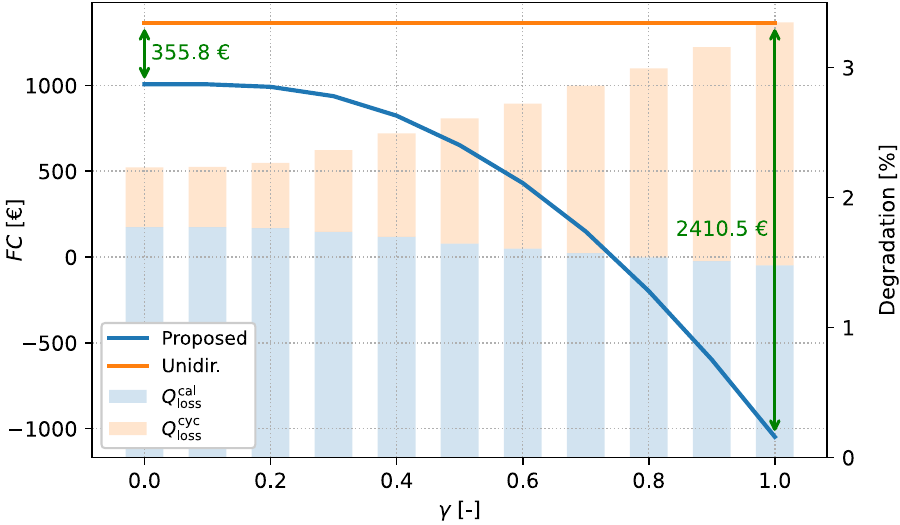}
    \caption{$\mathit{FC}$ for the proposed strategy and for unidirectional smart charging as a function of the V2G price ratio $\gamma$, for the VHGPV configuration. The stacked bars show the corresponding calendar and cycle degradation $Q_{\text{loss}}^{\mathrm{cal}}$ and $Q_{\text{loss}}^{\mathrm{cyc}}$ for the proposed strategy.}
    \label{fig:VHGPV_varingGamma_79kWh}
\end{figure}

The results in Table~\ref{tab:result_table} assume \(\gamma=1\), i.e., V2G exports paid at the full spot price, which is an optimistic scenario. However, different values of $\gamma$ can significantly affect energy revenue, and consequently the final cost. Fig.~\ref{fig:VHGPV_varingGamma_79kWh} shows, for the VHGPV configuration, $\mathit{FC}$ for the proposed bidirectional strategy and for the unidirectional smart charging as $\gamma$ varies from 0 to 1 with a step of 0.1, along with the corresponding calendar and cycle degradation for the proposed strategy.

As expected, the unidirectional case remains constant with respect to $\gamma$, since it does not perform V2G and is therefore independent of the V2G price ratio. For the proposed strategy, the point at $\gamma = 1$ coincides with the result reported in Table~\ref{tab:result_table}, yielding an economic gain of €2410.5 with respect to unidirectional smart charging.

As $\gamma$ decreases, V2G becomes progressively less attractive and the final cost $\mathit{FC}$ increases. For $\gamma$ between 0.7 and 0.8, $\mathit{FC}$ is approximately zero: in this range, the EV, supported by the PV system, fully offsets the entire electricity bill, corresponding to an economic gain of €1370.7 (i.e., the cost that would be incurred with unidirectional smart charging). For lower values of $\gamma$, $\mathit{FC}$ becomes positive, meaning that the user still reduces the annual electricity expenditure compared to the unidirectional case, but no net profit is generated. At $\gamma = 0$, V2G is never used because it provides no financial compensation; however, V2H is still exploited to increase the self-consumption, resulting in an economic gain of €355.8 compared to unidirectional smart charging.

The stacked bars in Fig.~\ref{fig:VHGPV_varingGamma_79kWh} illustrate the degradation composition for the proposed VHGPV configuration. Over the whole range of $\gamma$, the total degradation remains in a moderate interval of about 2.2–3.3\%. When $\gamma$ is small, aging is mainly calendar-driven, since the optimizer mostly exploits the EV for self-consumption and the battery spends longer periods parked at relatively high SoE. Considering the 2.1\% total degradation observed under unidirectional charging in Table~\ref{tab:result_table}, V2H-only operation provides economic benefit without significant additional battery wear. 
As $\gamma$ increases and V2G arbitrage becomes more attractive, the strategy induces higher energy throughput, increasing cycle aging while slightly reducing calendar aging due to a lower average SoE.

These results show that the proposed strategy yields economic benefits over the unidirectional smart charging across the entire range of $\gamma$ values. From the most favorable case with full V2G remuneration to the most conservative case without V2G compensation, the user always experiences a reduction in annual electricity expenditure, with additional degradation ranging from negligible ($\gamma=0$) to moderate ($\gamma=1$).

\subsection{Impact of Different Battery Capacity}
\begin{figure}[ht]
    \centering
    \includegraphics[width=0.9\linewidth]{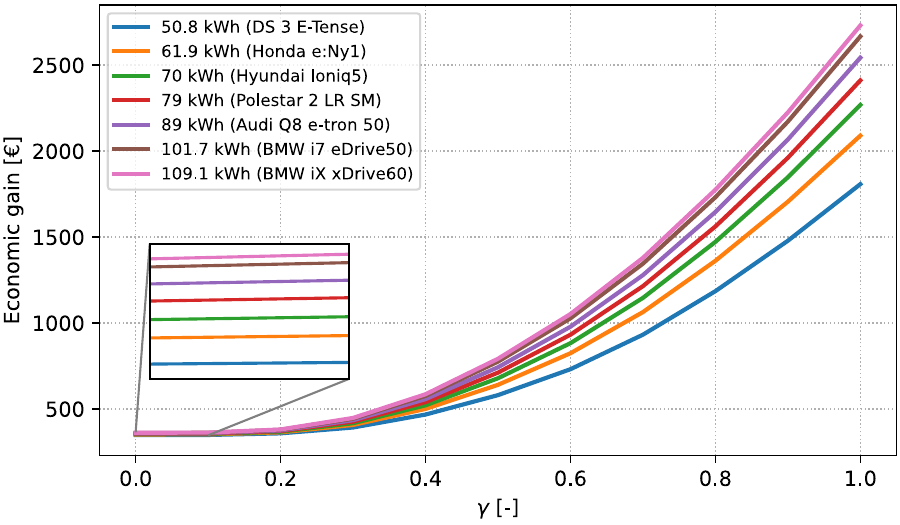}
    \caption{Economic gain of the proposed VHGPV strategy with respect to unidirectional smart charging for different EV battery capacities, as a function of the V2G price ratio $\gamma$.}
    \label{fig:VHGPV_EconGain_EVs}
\end{figure}

Additional simulations were conducted to evaluate the benefits of V2G and V2H operation for different battery capacities in the VHGPV configuration. 
Fig.~\ref{fig:VHGPV_EconGain_EVs} reports the economic gain, defined as the difference between the final cost of the unidirectional and the proposed strategy, for several EVs with battery capacities ranging from 50.8 to 109.1 kWh. 
The results show that, for any value of $\gamma$, a larger battery consistently yields higher benefits from bidirectional charging, and these benefits increase as $\gamma$ grows. In the best case, with $\gamma=1$, the annual economic gain ranges from approximately €1800 (50.8~kWh) up to approximately €2700 (109.1~kWh).

\begin{figure}[ht]
    \centering
    \includegraphics[width=0.9\linewidth]{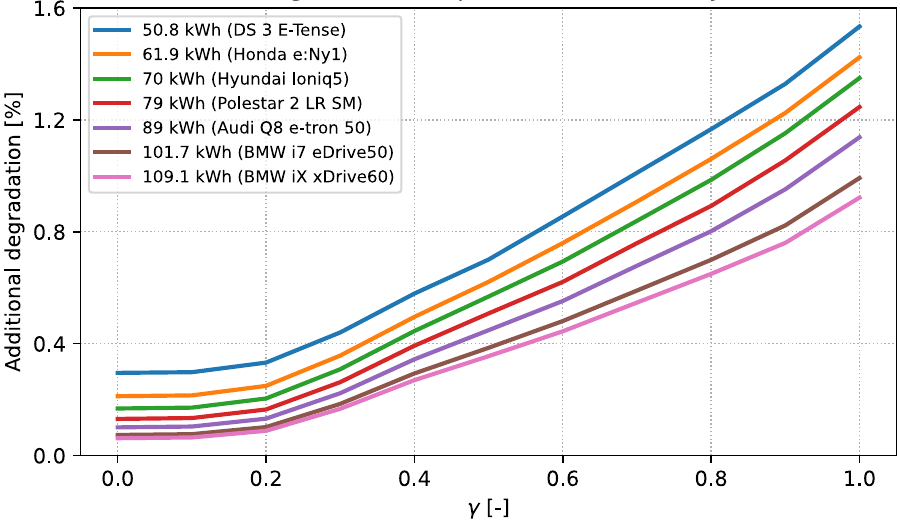}
    \caption{Additional battery degradation induced by bidirectional VHGPV operation, with respect to the unidirectional case, for different EV battery capacities, as a function of the V2G price ratio $\gamma$.}
    \label{fig:VHGPV_ExtraDegr_EVs}
\end{figure}

Fig.~\ref{fig:VHGPV_ExtraDegr_EVs} shows the additional annual degradation induced by bidirectional operation, computed as the difference between the degradation of the proposed bidirectional strategy and the unidirectional strategy. As expected, the additional degradation increases with $\gamma$, since a higher V2G price ratio leads to more energy being exchanged with the grid and thus to a larger number of charge/discharge events. 
The figure also shows a clear dependence on battery capacity: for any value of $\gamma$, the smallest pack (50.8~kWh) undergoes the largest additional degradation, while the largest pack (109.1~kWh) is affected the least. For the same exchanged energy, a smaller battery must operate with larger relative SoE swings and deeper cycles, which increase cycle-related aging; a larger pack delivers the same energy with shallower cycles and lower relative stress.

\begin{figure}[ht]
    \centering
    \includegraphics[width=0.9\linewidth]{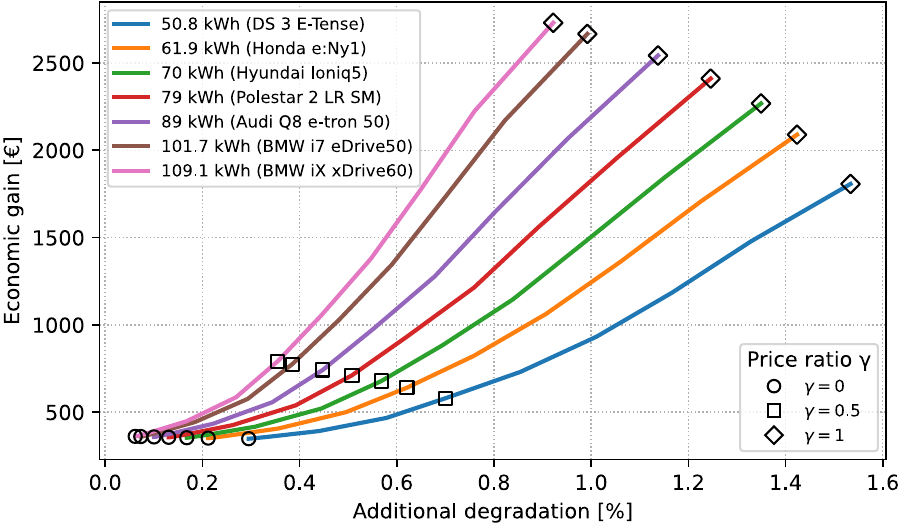}
    \caption{Economic gain of the proposed VHGPV strategy with respect to unidirectional smart charging as a function of the additional battery degradation, for different EV battery capacities and V2G price ratios $\gamma$.}
    \label{fig:VHGPV_EconGain_vs_ExtraDegr}
\end{figure}

Fig.~\ref{fig:VHGPV_EconGain_vs_ExtraDegr} summarizes the trade-off between economic gain and additional degradation. 
For all battery capacities, the economic gain increases monotonically with additional degradation, reflecting the fact that higher revenues are obtained by exploiting the battery more intensively. In the most favorable case $\gamma = 1$, the additional degradation ranges from 0.9\% (109.1~kWh) to 1.5\% (50.8~kWh), corresponding to economic gains of €2700 and €1800, respectively. 
Larger packs trace curves shifted upward and leftward, indicating higher monetary benefits per unit of additional degradation. 
This confirms that EVs equipped with larger batteries are more favorable for bidirectional services; however, smaller batteries still achieve gains of about €1800 at $\gamma=1$, with additional degradation not exceeding 1.55\%.

\subsection{Impact of Different Household Load} 

This section examines how the proposed strategy performs under different household load levels. The baseline load $\mathit{HL}$, introduced in Section~\ref{sec:2}, represents a standard apartment with average daily consumption of 21.6~kWh. To represent higher consumption, a scaled profile $\mathit{HL}\times4$ is constructed with average daily consumption of 86.4~kWh, corresponding to a large household. For this scenario, the Transformer-based forecaster is retrained on the scaled series to ensure consistent predictions.
In both cases, PV capacity remains at 10~kWh.

\begin{figure}[ht]
    \centering
    \includegraphics[width=0.9\linewidth]{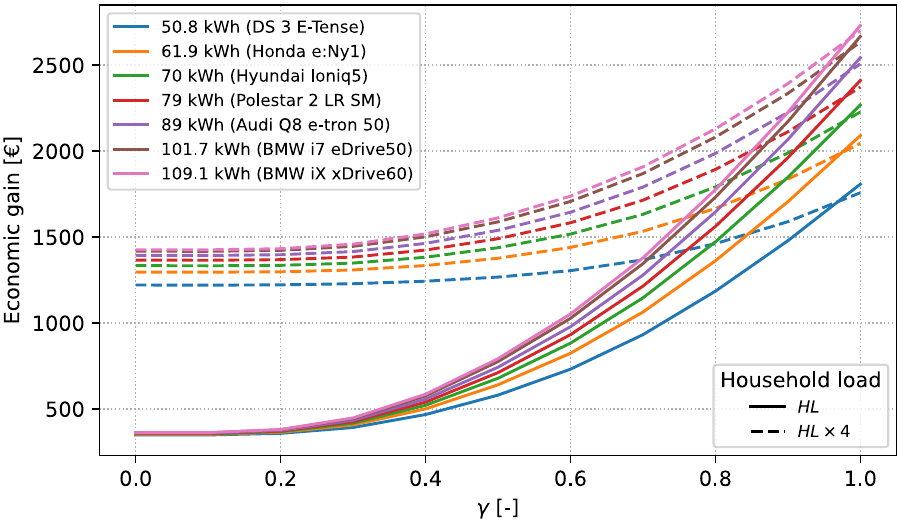}
    \caption{Economic gain of the proposed VHGPV strategy with respect to unidirectional smart charging, for two household load scenarios, as a function of the V2G price ratio $\gamma$.}
    \label{fig:VHGPV_4HL}
\end{figure}

Fig.~\ref{fig:VHGPV_4HL} compares the economic gain of the proposed strategy for the baseline $\mathit{HL}$ (solid lines) and scaled $\mathit{HL}\times 4$ (dashed lines) profiles. Increasing household demand shifts all curves upwards. 
For low-to-medium values of $\gamma$, economic gain is higher in the $\mathit{HL}\times 4$ case: larger demand provides more opportunities to substitute grid purchases via V2H when PV generation is insufficient.
Since V2H does not depend on $\gamma$, when $\gamma$ increases, the marginal gain for $\mathit{HL}\times 4$ grows slowly. In contrast, the standard $\mathit{HL}$ profile allows a larger share of battery energy to be used for V2G arbitrage, resulting in steeper economic gain as $\gamma$ rises.
At $\gamma = 1$, $\mathit{HL}$ achieves higher gain than $\mathit{HL}\times 4$, as the EV can exploit more profitable V2G arbitrage with the grid rather than being constrained to supply household demand. 
In both load scenarios, larger battery capacity consistently yields higher economic gain.

\subsection{Impact of EV Pickup Time Uncertainty}
In all previous results, it is assumed that when the user parks the EV, they declare a pickup time, assumed to be accurate. The optimization then spans this interval, and the planned charging/discharging schedule is executed under the implicit assumption that the EV remains connected until the declared departure time.
In practice, however, the actual pickup time may deviate from the announced one: the user may retrieve the EV earlier (early pickup) or later (late pickup) than expected. Early pickup truncates the planned trajectory, potentially reducing the economic benefit and altering the desired SoE at the pickup time. In the late-pickup case, the EV may remain idle for additional hours, leaving V2H/V2G capacity unexploited.

To assess the impact of this uncertainty in the VHGPV setup, we perturb the declared departure times. For each parking interval, the nominal pickup time is randomly shifted earlier or later by up to a given percentage 
$e \in \{10,30,50\}\,\%$ of the parking duration. A change is applied with probability $p_{\mathrm{change}} = 0.8$, to ensure sufficient variation. This procedure generates scenarios with both early and late pickups, allowing us to quantify the impact of increasing timing uncertainty on the final cost.
\begin{figure}[ht]
    \centering
    \includegraphics[width=0.9\linewidth]{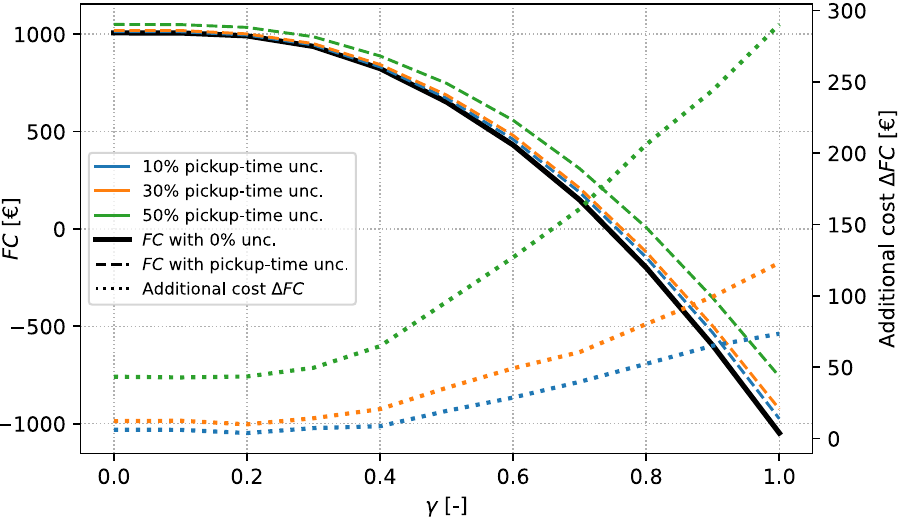}
    \caption{Impact of pickup time uncertainty in the VHGPV case (79~kWh EV): final cost $\mathit{FC}$ and additional cost $\Delta \mathit{FC}$ relative to the no-uncertainty baseline versus the V2G price ratio $\gamma$.}
    \label{fig:pickupUncert}
\end{figure}
Fig.~\ref{fig:pickupUncert} reports the final cost $\mathit{FC}$ and the additional cost $\Delta \mathit{FC}$ relative to the no-uncertainty baseline, as functions of $\gamma$ and for different pickup time uncertainty levels.

At $\gamma = 0$, uncertainty has limited impact: even 50\% deviation adds less than €50 to annual cost. Conversely, for $\gamma = 1$ sensitivity increases substantially, with $\Delta FC$ rising from approximately €75 (10\% uncertainty) to nearly €300 (50\% uncertainty). This shows that pickup time uncertainty is relatively benign when V2G remuneration is low, whereas under high V2G price ratios the economic performance becomes significantly more sensitive to deviations from the declared departure time, especially for large pickup time deviations. 
Nonetheless, even in the worst-case scenario, the additional cost does not overturn the economic benefit of the proposed strategy, whose annual cost remains clearly favorable compared to the unidirectional smart charging.

\subsection{Impact of PV System Sizing}
\begin{figure}[ht]
    \centering
    \includegraphics[width=0.9\linewidth]{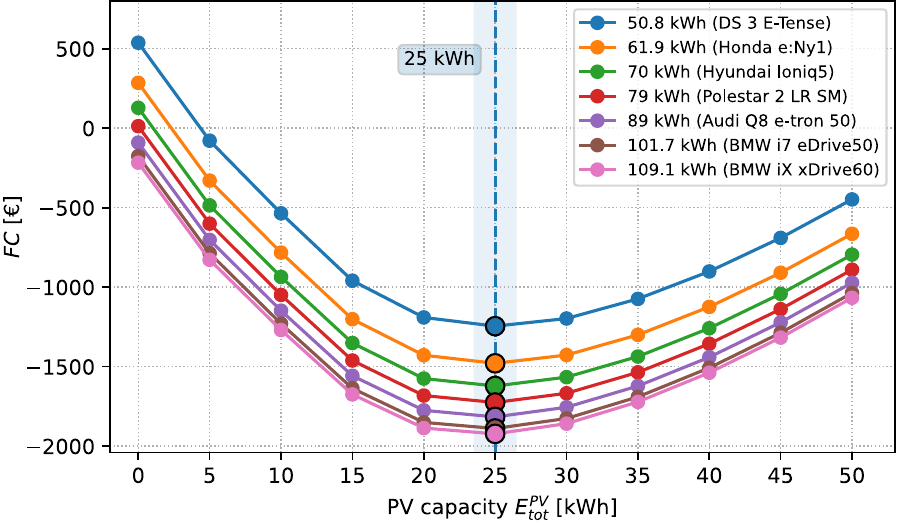}
    \caption{Sensitivity of the final cost $\mathit{FC}$ to the PV capacity in the VHGPV configuration for different EV battery capacities at $\gamma = 1$.}
    \label{fig:PVsize}
\end{figure}

Previous results assume a fixed PV system size of 10~kWh, as specified in Section~\ref{sec:5}. To assess how the PV sizing affects performance, a sensitivity analysis is carried out by varying the total installed PV capacity from 0 to 50~kWh in steps of 5~kWh, with $\gamma = 1$. Fig.~\ref{fig:PVsize} shows the resulting annual final cost $\mathit{FC}$ as a function of PV capacity for all considered EV battery sizes.

Increasing PV capacity raises investment cost but also increases revenue from self-consumption and grid export. 
For each EV, an economically optimal PV size exists around 25~kWh, where $\mathit{FC}$ attains its minimum value, with lower absolute costs achieved by EVs equipped with larger batteries. 
Beyond this point, further enlarging the PV size does not translate into additional exploitable PV energy because of the imposed upper bound $E_{\mathrm{PV},\max} = 11$~kWh; the investment cost continues to grow while exploitable PV energy saturates, causing $\mathit{FC}$ to rise.
However, a 25~kWh system would require approximately 100~m$^2$ of rooftop area, at the upper limit of what is feasible for single-family dwellings \cite{sizePV_SE}. 

\section{Conclusion}
This paper presented an online, aging-aware energy management strategy based on a shrinking-horizon MPC formulation with a linearized battery degradation model and a Transformer-based forecaster for household load and solar irradiance. The controller jointly optimizes power flows between the EV, household, grid, and rooftop PV, exploiting V2G and V2H flexibility for arbitrage and self-consumption.

A full-year case study under realistic Swedish conditions shows that the proposed strategy attains the lowest annual cost among all evaluated approaches.
Relative to smart unidirectional charging, bidirectional operation with rooftop PV provides an economic gain of up to €2410.5 per year under favorable V2G pricing, at the cost of only 1.27\% additional battery degradation. Even without V2G remuneration, V2H alone delivers €355.8 annual savings with negligible extra degradation. 
Compared to operating without PV, adding a 10 kWh rooftop PV system to the bidirectional strategy increases annual profit by €1060.7.
Sensitivity analyzes over V2G price ratio, battery capacity, household demand, and pickup time uncertainty confirm that these benefits persist across a wide range of scenarios, with larger batteries proving particularly well-suited for bidirectional services.

Overall, these results indicate that, under the assumptions adopted in this study and across a wide range of operating conditions, enabling bidirectional charging improves the user’s economic outcome while keeping battery degradation within acceptable levels.
Future work will extend the framework to multi-user and distribution-network-aware settings to assess how bidirectional charging can support grid resilience under both normal operation and fault conditions.

\bibliographystyle{IEEEtran} 
\bibliography{Section/bib} 

@misc{2025IEAGlobalEVOutlook,
  author       = {{International Energy Agency}},
  title        = {{Global EV Outlook 2025}},
  year         = {2025},
  publisher    = {IEA},
  address      = {Paris},
  url          = {https://www.iea.org/reports/global-ev-outlook-2025},
  note         = {Licence: CC BY 4.0}
}

@misc{IEA2024WorldEnergyInvestment,
  author       = {{International Energy {Agency}}},
  title        = {{World Energy Investment 2024}},
  year         = {2024},
  publisher    = {IEA},
  address      = {Paris},
  url          = {https://www.iea.org/reports/world-energy-investment-2024},
  note         = {Licence: CC BY 4.0}
}

@Article{2019v2xReview,
AUTHOR = {Vadi, Seyfettin and Bayindir, Ramazan and Colak, Alperen Mustafa and Hossain, Eklas},
TITLE = {A Review on Communication Standards and Charging Topologies of {V2G and V2H} Operation Strategies},
JOURNAL = {Energies},
VOLUME = {12},
YEAR = {2019},
NUMBER = {19},
ARTICLE-NUMBER = {3748},
ISSN = {1996-1073},
DOI = {10.3390/en12193748}
}

@ARTICLE{2015v2hv2g,
  author={Erdinc, Ozan and Paterakis, Nikolaos G. and Mendes, Tiago D. P. and Bakirtzis, Anastasios G. and P. S. Catalão, João},
  journal={IEEE Trans. Smart Grid}, 
  title={Smart Household Operation Considering Bi-Directional {EV} and {ESS} Utilization by Real-Time Pricing-Based {DR}}, 
  year={2015},
  volume={6},
  number={3},
  pages={1281-1291},
  doi={10.1109/TSG.2014.2352650}}

@ARTICLE{2023MARLv2g,
  author={Dong, Jiawei and Yassine, Abdulsalam and Armitage, Andy and Hossain, M. Shamim},
  journal={IEEE Trans. Intell. Transp. Syst.}, 
  title={Multi-Agent Reinforcement Learning for Intelligent {V2G} Integration in Future Transportation Systems}, 
  year={2023},
  volume={24},
  number={12},
  pages={15974-15983}}

@article{2022v2hv2gGermany,
title = {Revenue opportunities by integrating combined vehicle-to-home and vehicle-to-grid applications in smart homes},
journal = {Appl. Energy},
volume = {307},
pages = {118187},
year = {2022},
issn = {0306-2619},
doi = {https://doi.org/10.1016/j.apenergy.2021.118187},
author = {Timo Kern and Patrick Dossow and Elena Morlock}
}

@article{2022onlineV2GPengfei,
title = {Online battery-protective vehicle to grid behavior management},
journal = {Energy},
volume = {243},
pages = {123083},
year = {2022},
issn = {0360-5442},
doi = {https://doi.org/10.1016/j.energy.2021.123083},
author = {Shuangqi Li and Pengfei Zhao and Chenghong Gu and Da Huo and Xianwu Zeng and Xiaoze Pei and Shuang Cheng and Jianwei Li}
}

@Article{2024OnlineV2GPSO,
AUTHOR = {Zhang, Qingguang and Ikram, Mubasher and Xu, Kun},
TITLE = {Online Optimization of Vehicle-to-Grid Scheduling to Mitigate Battery Aging},
JOURNAL = {Energies},
VOLUME = {17},
YEAR = {2024},
NUMBER = {7},
ARTICLE-NUMBER = {1681},
ISSN = {1996-1073},
DOI = {10.3390/en17071681}
}

@Article{2019JapanEVv2hPV,
AUTHOR = {Kataoka, Ryosuke and Shichi, Akira and Yamada, Hiroyuki and Iwafune, Yumiko and Ogimoto, Kazuhiko},
TITLE = {{Comparison of the Economic and Environmental Performance of V2H and Residential Stationary Battery: Development of a Multi-Objective Optimization Method for Homes of EV Owners}},
JOURNAL = {World Electr. Veh. J.},
VOLUME = {10},
YEAR = {2019},
NUMBER = {4},
ARTICLE-NUMBER = {78},
ISSN = {2032-6653},
DOI = {10.3390/wevj10040078}
}

@article{2025PVoptimalsizingAustralia,
author = {Azarbakhsh, Golsa and Mahmoudi, Amin and Kahourzade, Solmaz and Yazdani, Amirmehdi and Mahmud, Apel},
title = {Optimal Sizing of Photovoltaic and Battery Energy Storage for Residential Houses in South Australia by Considering Vehicle-to-Home Operation},
journal = {IET Renew. Power Gener.},
volume = {19},
number = {1},
pages = {e70053},
doi = {https://doi.org/10.1049/rpg2.70053},
year = {2025}
}

@Article{2025NNMPCv2gPV,
AUTHOR = {Dankar, Ossama and Tarnini, Mohamad and El Ghaly, Abdallah and Moubayed, Nazih and Chahine, Khaled},
TITLE = {{A Neural Network-Based Model Predictive Control for a Grid-Connected Photovoltaic–Battery System with Vehicle-to-Grid and Grid-to-Vehicle Operations}},
JOURNAL = {Electricity},
VOLUME = {6},
YEAR = {2025},
NUMBER = {2},
ARTICLE-NUMBER = {32},
ISSN = {2673-4826},
DOI = {10.3390/electricity6020032}
}

@ARTICLE{2019EVchargingPV,
  author={Yan, Qin and Zhang, Bei and Kezunovic, Mladen},
  journal={IEEE Trans. Smart Grid}, 
  title={{Optimized Operational Cost Reduction for an EV Charging Station Integrated With Battery Energy Storage and PV Generation}}, 
  year={2019},
  volume={10},
  number={2},
  pages={2096-2106},
  doi={10.1109/TSG.2017.2788440}}

@book{park2019_economics,
  author   = {Park, Chan},
  title    = {{Fundamentals of Engineering Economics}},
  publisher = {Upper Saddle River, NJ, USA: Pearson Education},
  year     = {2013}
}

@ARTICLE{2025mpcV2Grangeanxiety,
  author={Lu, Chuan-Fan and Liu, Guo-Ping and Yu, Yi and Cui, Jinqiang},
  journal={IEEE Trans.  Transp. Electr.}, 
  title={A Coordinated Model Predictive Control-Based Approach for Vehicle-to-Grid Scheduling Considering Range Anxiety and Battery Degradation}, 
  year={2025},
  volume={11},
  number={2},
  pages={5688-5699},
  doi={10.1109/TTE.2024.3488075}}

@misc{IEA_PVPS_Sweden_2023,
  author       = {{IEA PVPS Task 1}},
  title        = {National Survey Report of {PV} Power Applications in Sweden 2023},
  year         = {2024},
  month        = apr,
  howpublished = {\url{https://iea-pvps.org/national_survey/national-survey-report-of-pv-power-applications-in-sweden-2023/}},
  note         = {Accessed: 2025-10-17}
}

@inproceedings{2017TransformerGoogle,
author = {Vaswani, Ashish and Shazeer, Noam and Parmar, Niki and Uszkoreit, Jakob and Jones, Llion and Gomez, Aidan N. and Kaiser, \L{}ukasz and Polosukhin, Illia},
title = {Attention is all you need},
year = {2017},
booktitle = {Proc. 31st Int. Conf. Neural Inf. Process. Syst.},
pages = {6000–6010},
numpages = {11},
}

@article{2024TransformerLoadForecasting,
title = {A Transformer based approach to electricity load forecasting},
journal = {Electr. J.},
volume = {37},
number = {2},
pages = {107370},
year = {2024},
issn = {1040-6190},
doi = {https://doi.org/10.1016/j.tej.2024.107370},
author = {Jun Wei Chan and Chai Kiat Yeo}
}

@misc{2021_hl_data, 
    title = {End-Use Load Profiles for the {U.S.} Building Stock [data set]}, 
    author = {{National Renewable Energy Laboratory (NREL)}},  
    doi = {https://dx.doi.org/10.25984/1876417}, 
    url = {https://data.openei.org/submissions/4520}, 
    place = {United States}, 
    year = {2021}
}

@misc{2021_sun_data_goteborg,
	doi={10.5878/a2h2-4s63},
	language={en},
	publisher={University of Gothenburg},
	title={{Urban climate data for Gothenburg, 1983-2020}},
	url={https://doi.org/10.5878/a2h2-4s63},
	author={Rayner, David and Lindberg, Fredrik and Kukulies, Julia and Thorsson, Sofia and Wallenberg, Nils},
	date=2021,
	year=2021,
}

@article{2023batteryNature,
  title={Electric vehicle batteries alone could satisfy short-term grid storage demand by as early as 2030},
  author={Xu, Chengjian and Behrens, Paul and Gasper, Paul and Smith, Kandler and Hu, Mingming and Tukker, Arnold and Steubing, Bernhard},
  journal={Nature Commun.},
  volume={14},
  number={1},
  pages={119},
  year={2023},
  publisher={Nature Publishing Group UK London}
}

@article{2025YukiDrivingPatterns,
title = {{Assessment of real-world driving patterns for electric vehicles: an on-board measurements study from Sweden}},
journal = {Appl. Energy},
volume = {401},
pages = {126608},
year = {2025},
issn = {0306-2619},
doi = {https://doi.org/10.1016/j.apenergy.2025.126608},
author = {Yuki Kobayashi and Maria Taljegard and Filip Johnsson}
}

@misc{trafikanalys2024,
  author       = "{Trafikanalys}",
  title        = "{Körsträckor 2024}",
  howpublished = "\url{https://www.trafa.se/vagtrafik/korstrackor/}",
  note         = "Accessed: Oct. 29, 2025"
}

@misc{goteborgenergi,
  author       = {{Göteborg Energi}},
  title        = {{Göteborg Energi Official Website}},
  howpublished = {\url{https://www.goteborgenergi.se/}},
  note         = {Accessed: Nov. 4, 2025}
}

@misc{nordpool,
  author       = {{Nord Pool}},
  title        = {{Nord Pool Power Market -- Official Website}},
  howpublished = {\url{https://www.nordpoolgroup.com/}},
  note         = {Accessed: Nov. 4, 2025}
}

@misc{EVDatabase,
  title        = {{EV Database: All Electric Vehicles}},
  howpublished = {\url{https://ev-database.org/}},
  note         = {{Accessed: 2025-11-17}}
}

@misc{battCost_data,
  author       = {{U.S. Dept. of Energy}},
  title        = {{Electric Vehicle Battery Pack Costs for a Light-Duty Vehicle in 2023 Are 90\% Lower than in 2008, according to DOE Estimates}},
  howpublished = {https://www.energy.gov/eere/vehicles/articles/fotw-1354-august-5-2024-electric-vehicle-battery-pack-costs-light-duty},
  year         = {2024}  
}

@misc{battCost_data_1,
  author       = {{BloombergNEF}},
  title        = {{Lithium-Ion Battery Pack Prices See Largest Drop Since 2017, Falling to \$115 per Kilowatt-Hour}},
  howpublished = {https://about.bnef.com/insights/commodities/lithium-ion-battery-pack-prices-see-largest-drop-since-2017-falling-to-115-per-kilowatt-hour-bloombergnef},
  note         = {Accessed: 2025-11-18},
  institution  = {BloombergNEF},
}

@misc{sizePV_SE,
  author    = {Otovo},
  title     = {Hur mycket solceller behöver man?},
  year      = {2023},
  url       = {https://www.otovo.se/blog/solpaneler-solceller/hur-mycket-solceller-behover-man/},
  note      = {Accessed: 2025-11-18}
}

@Article{casadi,
  Author = {Joel A E Andersson and Joris Gillis and Greg Horn
            and James B Rawlings and Moritz Diehl},
  Title = {{CasADi} -- {A} software framework for nonlinear optimization
           and optimal control},
  Journal = {Math. Program. Comput.},
  Year = {2018},
}

@ARTICLE{Qingbo_EVcons,
  author={Zhu, Qingbo and Huang, Yicun and Feng Lee, Chih and Liu, Peng and Zhang, Jin and Wik, Torsten},
  journal={IEEE Trans. Transp. Electr.}, 
  title={Predicting Electric Vehicle Energy Consumption From Field Data Using Machine Learning}, 
  year={2025},
  volume={11},
  number={1},
  pages={2120-2132},
  doi={10.1109/TTE.2024.3416532}}

\vfill

\end{document}